\documentstyle[aps]{revtex}
\newcommand{\ds}{\displaystyle}
\newcommand{\dsf}{\ds\frac}

\newcommand{\beq}{\begin{equation}}
\newcommand{\eeq}{\end{equation}}

\large

\begin{document}
\large

\begin{center}
\Large\bf
Running Waves in the Mixed State of Type II Superconductors
\vskip 0.1cm
{\normalsize\bf N.A.\,Taylanov, G.B.\, Berdiyorov}\\
\vskip 0.1cm
{\large\em Theoretical Physics Department,
Institute of Applied Physics,\\
National University of Uzbekistan,\\
E-mail: taylanov@iaph.tkt.uz}
\end{center}
\begin{center}
{\bf Abstract}
\end{center}

\begin{center}
\mbox{\parbox{14cm}{\small
         Nonlinear dynamics of the thermal and electromagnetic
instabilities of the mixed state in type II superconductors has been
analysed taking into account the effect of dissipation and dispersion.
The existance of nonlinear running waves describing the final stage of
evolution of the thermomagnetic instability in superconductors
is demonstrated analitically.
}}
\end{center}
\vskip 0.5cm

         The evolution of the thermal $(T)$ and electromagnetic
$(\vec E, \vec H)$ perturbations is described by the nonlinear equation
of the thermal conductivity [1]
\beq
\nu\frac{dT}{dt}=\nabla [\kappa\nabla T]+\vec j\vec E,
\eeq
by the Maxwell equations
\beq
rot\vec E=-\dsf{1}{c}\dsf{d\vec H}{dt},
\eeq
\beq
rot{\vec H}=\dsf{4\pi}{c}\vec j
\eeq
and by the equation of the resistive state
\beq
\vec j=\vec j_{c}(T,\vec H)+\vec j_{r}(\vec E),
\eeq
where $\nu=\nu(T)$ and $\kappa=\kappa(T)$ are the heat capacity and the
thermal conductivity respectively; $\vec j_c$ is the critical current density
and $\vec j_r$ is the resistance current density.

        The above system is essentially nonlinear because the right-hand
part of Eq.(1) contains a term describing the Joule heat evolution
in the region of the resistive phase.
Such a set (1)-(4) of nonlinear parabolic differential equations in
partial derivatives has no exact analytical solution.

        Let us consider a planar semi-infinite sample $(x>0)$ placed in
external magnetic field $\vec H=(0, 0, H_{e})$ growing at a constant rate
$\dsf{d\vec H}{dt}=const$. According to the Maxwell equation (2), there is a vortex
electric field $\vec E=(0, E_e, 0)$ in the sample, directed parallel to the
current density $\vec j$: $\vec E\parallel \vec j$; where $H_e$ is the
amplitude of the external magnetic field and $E_e$ is the amplitude of the
external electric field.

        The solution of the system of equations (1)-(4) may be
presented as a function of new automodel variable $\xi(x,t)$:
\beq
\begin{array}{l}
T=T[\xi(x,t)],\\
\quad\\
E=E[\xi(x,t)],\\
\quad\\
j=j[\xi(x,t)].\\
\end{array}
\eeq

        Substituting (5) into the system (1)-(4) gives, as a result of simple
differentation, the following system

\beq
\dsf{d\xi}{dt}\left[\nu\dsf{dT}{d\xi}\right]=
\kappa\left\{\dsf{d^2\xi}{dx^2}\dsf{dT}{d\xi}+
\left(\dsf{d\xi}{dx}\right)^2
\dsf{d^2T}{d\xi^2}\right\}+[j_c(T)+j_r(E)]E\,,
\eeq

\beq
\dsf{d^2\xi}{dx^2}\dsf{dE}{d\xi}+\left(\dsf{d\xi}{dx}\right)^2
\dsf{d^2E}{d\xi^2}=\dsf{4\pi}{c^2}
\left[\dsf{dj_c}{dT}\dsf{dT}{d\xi}+\dsf{dj_r}{dE}\dsf{dE}{d\xi}\right]
\dsf{d\xi}{dt}\,.
\eeq

       In order the system (6),(7) was only function from $\xi$
 at the substation (5) it is required carrying out the following conditions:
\beq
\dsf{d\xi}{dt}=A(\xi)\,,
\eeq
\beq
\left(\dsf{d\xi}{dx}\right)^2=B(\xi)\dsf{d\xi}{dt}=G(\xi)\,,
\eeq
\beq
\dsf{d^2\xi}{dx^2}=C(\xi)\dsf{d\xi}{dt}\,,
\eeq
where $A,C,G$ are functions from $\xi$, the type of which will be determined
below. Solving first two system of equations (8)-(10) we have a relation
\beq
G(\xi)\dsf{dA}{d\xi}=A(\xi)\dsf{dG}{d\xi}\,.
\eeq

Whence just fllows relationship between $G$ and $A$
\beq
G(\xi)=\dsf{1}{u}A(\xi)\,,
\eeq
where $u$  is a free constant of integrating of the equation (11). From
(8) and (9) follows that $\xi(x,t)$ must satisfy single-line equation
in private derivation
\beq
\dsf{d\xi}{dt}=u\dsf{d\xi}{dx}
\eeq
the only solution of which is the function
\beq
\xi(x,t)=F(x-ut)\,.
\eeq

        Using (14) we can immediately obtain

\beq
G(\xi)=1,\quad A(\xi)=-Fu,\quad C(\xi)=0.
\eeq

It is possible to ensure $F=1$ by the transformation of coordinates and time.
Thereby, we find final automodel substitution
\beq
\xi=x-ut,
\eeq

corresponding to solution of running type wave [2].

        For the automodelling solution of the form (16), describing
a running wave moving at a constant velocity $v$ along the $x$ axis,
the system of equations (1)-(4) takes the following form
\beq
- v\left[N(T)-N(T_0)\right]=\kappa\dsf{dT}{d\xi}-\frac{c^2}{4\pi v}E^2,
\eeq
\beq
\dsf{dE}{d\xi}=-\dsf{4\pi v}{c^2}j,
\eeq
\beq
E=\dsf{v}{c}H.
\eeq

       The thermal and electrodynamic boundary conditinos for equations
(17)-(19) are as follows:
\beq
\begin{array}{l}
T(\xi\rightarrow+\infty)=T_0, \dsf{dT}{d\xi}(\xi\rightarrow-\infty)=0,\\
\quad\\
E(\xi\rightarrow+\infty)=0,   E(\xi\rightarrow-\infty)=E_e,\\
\end{array}
\eeq
where $T_0$ is the temperature of the cooling medium.

          Let us consider the Bean-London model of the critical state for
the dependence $j_{c}(T,H)$ [3]
\beq
j_{c}(T)=j_0[1-a(T-T_{0})]
\eeq
where $j_{0}$ is the equilibtium current density,
$a$ is the thermal haet softening coefficient of the magnetic flux pinning
force.
        The characteristical field dependence of $j_r(E)$ in the region of
sufficiently strong electric field $(E>E_f)$ can be aproximated by the
piecewise linear function $j_r\approx\sigma_f E$, where
$\sigma_f=\dsf{\eta c^2}{H\Phi_0}\approx \sigma_n H_{c_2}/H$ is the
effictive conductivity in the flux flow regime; $\eta$ is the viscous
coefficient,$\Phi_0=\dsf{\pi h c}{2e}$ is the magnetic
flux quantum, $\sigma_n$ is the conductivity in the normal state;
$E_f$ is the boundary of the linear area in the voltage-current
characteristics of the sample.

     Excluding variables $T(\xi)$ and $H(\xi)$ from Eqs.(17) and (19) ,
and taking into account the boundary conditions (20), we obtain an equation
describing the electric field $E(\xi)$ distribution ($E$-wave):
\beq
\dsf{d^2 E}{d\xi^2}+\left[\dsf{4\pi v}{c^2}\dsf{dj_r}{dE}\dsf{dE}{d\xi}+
\dsf{4\pi v^2a}{c^2}\dsf{N(T)-N(T_0)}{\kappa (T)}\right]-
\dsf{aE^2}{2\kappa (T)}=0,
\eeq
where the dependency $T=T\left(E,\dsf{dE}{d\xi}\right)$ is defined by
expression (2), (4) and have the form

\beq
T=T\left(E,\dsf{dE}{d\xi}\right)=T_0+\dsf{1}{a}\left[j_0+j_r(E)+
\dsf{4\pi v}{c^2}\dsf{dE}{d\xi}\right].
\eeq

Here $N(T)=\int\limits_{0}^{T}\nu(T)dT$.

The analysis of the phase plane ($E,\dsf{dE}{d\xi}$) of Eq.(22) shows that
there are two equilibrium points: $E_0=0$, $T=T_0$ is the stabile node and
$E=E_e$, $T=T*=T(E_e,0)$ is the saddle. The solution may be representedof
in the form of the shock-wave-type with amplitude $E_e$, by joining these
two equilibrium points.
The velocity of  $E$-wave is determined by the Eq. (22) with account
of the boundary conditions (20):
\beq
v_{E}^{2}=\dsf{c^2}{8\pi}\dsf{E_{e}^{2}}
{N\left[T_0+\dsf{1}{a}[j_c(T)+j_r(E)]\right]-N(T_0)}.
\eeq
Using Eqs. (19) and (22) we find the expression for distribution of the
magnetic field $H$ in the case of $H$-wave :
\beq
\dsf{d^2H}{d\xi^2}+\dsf{4\pi v}{c^2}\left[\left.\dsf{dj_r}{dE}
\right|_{E=\dsf{v}{c}H}\dsf{dH}{d\xi}+ca\dsf{N(T)-N(T_0)}
{\kappa (T)}\right]-\dsf{av}{2c}\dsf{H^2}
{\kappa (T)}=0.
\eeq

The velocity of $H$ wave is connected with its amplitude by the following expression

\beq
N(T)-N(T_0)=\dsf{H_{e}^{2}}{8\pi}.
\eeq
The condition (26) reflects the adiabatical character of the wave propagation:
the magnetic field energy transferred by the wave ensures local heating of
the sample in the close vicinity of the wave front.
Thereby, depending on the external conditions at the sample surfase
there may exist two kinds of thermomagnetic waves.

The system of Eqs. (1)-(4) is invariant with respect to an arbitrary translation.
Therefore, the wave propagation conditions can be found for an arbitrary
critical current density depending on $T$ and $H$. The results can also be
obtained for an arbitrary temperature dependence of thermophysical parameters
$\nu$ and $\kappa$ of superconducting material and for an arbitrary function
$j_r(E)$.

\newpage
\centerline{\large \bf NIZAM A.TAYLANOV}
\begin{tabbing}
{\bf Address:} \\
Theoretical Physics Department,\\
Institute of Applied Physics,\\
National University of Uzbekistan,\\
Vuzgorodok, 700174, Tashkent, Uzbekistan\\
Telephone:(9-98712)-461-573\\
fax: (9-9871)-144-77-28\\
e-mail: taylanov@iaph.tkt.uz \\
\end{tabbing}
\end{document}